\documentclass[
unsortedaddress,
 amsmath,amssymb,
 aps,
pre,
]{revtex4-2}
\usepackage{amsmath}
\usepackage{amssymb}
\usepackage{bm}
\usepackage{mathrsfs}
\usepackage{graphicx}

\begin{document}



\title{Thread-safe multiphase lattice Boltzmann model for droplet and bubble dynamics at high density and viscosity contrasts}

\author{Marco Lauricella}
\affiliation{%
 Istituto per le Applicazioni del Calcolo, Consiglio Nazionale delle Ricerche,Via dei Taurini 19,Rome,00185, Italy
}%
\author{Adriano Tiribocchi}%
\affiliation{%
 Istituto per le Applicazioni del Calcolo, Consiglio Nazionale delle Ricerche,Via dei Taurini 19,Rome,00185, Italy
}%
\affiliation{INFN "Tor Vergata",Via della ricerca scientifica 1,Rome,00133, Italy}  
            
\author{Sauro Succi}
\affiliation{Center for Life Nano- \& Neuro-Science, Fondazione Istituto Italiano di Tecnologia,
            Viale Regina Elena 291, 
            Rome,
            00161, 
            Italy
}%
\author{Luca Brandt}
\affiliation{Dipartimento di Ingegneria dell’Ambiente, del Territorio e delle Infrastrutture, Politecnico di Torino, Torino, Italia
}%
\author{Aritra Mukherjee}
\affiliation{%
 Department of Energy and Process Technology, NTNU,Norway
}%
\author{Michele La Rocca}
\affiliation{Department of Civil, Computer Science and Aeronautical Technologies Engineering, Roma Tre University,Via Vito Volterra,Rome, 00146, Italy
}%
\author{Andrea Montessori}
\email{andrea.montessori@uniroma3.it}
\affiliation{Department of Civil, Computer Science and Aeronautical Technologies Engineering, Roma Tre University,Via Vito Volterra,Rome, 00146, Italy
}%

\begin{abstract}
This study presents a high-order, thread-safe version of the lattice Boltzmann (LBM) method, incorporating an interface-capturing equation, based on the conservative Allen-Cahn equation, to simulate incompressible two-component systems with high-density and viscosity contrasts. The method utilizes a recently proposed thread-safe implementation optimized for shared memory architectures and it is employed to reproduce the dynamics of droplets and bubbles on several test cases with results in agreement with experiments and other numerical simulations from the literature. The proposed approach offers promising opportunities for high-performance computing simulations of realistic fluid systems with high-density and viscosity contrasts for advanced applications in environmental, atmospheric and meteorological flows, all the way down to microfluidic and biological systems, particularly on GPU-based architectures.
\end{abstract}

\maketitle




\section{Introduction}

Simulating multi-component systems with high density and viscosity contrasts is crucial across a wide range of applications, including optimizing industrial processes \cite{worner2012numerical,ho2013multiscale}  investigating environmental, atmospheric and meteorological flows \cite{parker1989multiphase, grabowski2013growth} and advancing our knowledge over microfluidic and biological systems \cite{bogdan2022stochastic, guo2012droplet}. These examples highlight the need for a comprehensive set of accurate theoretical and numerical models capable of faithfully reproducing the physics of multicomponent and multiphase flows, especially those characterized by high density and viscosity ratios.

However, modeling the physics of fluids with interfaces is a formidable challenge due to the need of accurately capturing the interplay among forces of  different natures and operating at strikingly different spatial and temporal scales. These include near-contact interactions \cite{montessori2019mesoscale}, capillary, inertial, and drag forces \cite{michaelides2022multiphase}, which span a broad spectrum of multi-scale physical phenomena. Examples range from turbulence-interface interactions \cite{constante2021direct} to capillarity effects \cite{montessori2019jetting,montessori2018elucidating}, turbulent emulsification-induced complex rheology \cite{yi2023recent}, and heat transfer with phase change \cite{martinez2021new}.
The inherently multi-scale nature of multiphase flows, combined with the complex physics associated with dynamic interfacial rearrangements, poses significant challenges in unveiling the multitude of physical mechanisms driving these systems' space-time evolution. As a result, understanding the fundamental physics of multicomponent systems with density and viscosity contrasts necessitates the development of accurate, high-fidelity, and efficient numerical models.
From a computational perspective, the primary challenge lies in developing high-performance computing codes that can:
i) span temporal and spatial scales necessary for capturing long-timescale behavior of fluid systems and
ii) resolve the intricate dynamic interactions between turbulence and fluid interfaces \cite{exasc2,exasc1}.
Over the past two decades, the lattice Boltzmann method \cite{kruger2017lattice,montessori2018lattice} has gained significant traction in the computational fluid dynamics (CFD) community as a hydrodynamic solver in kinetic form. Its popularity stems from its conceptual and practical simplicity, as well as its intrinsic efficiency \cite{montessori2018lattice, exasc1}. Unlike traditional hydrodynamic methods, where the convective derivative is non-linear and non-local, the LBM separates non-linearity and non-locality into two distinct operators: the collision step (local and non-linear) and the streaming step (linear, non-local, and exact to machine precision).
Additionally, LBM algorithms are simpler to implement than traditional CFD methods, relying on straightforward, local update rules that greatly simplify coding and debugging. The local nature of the collision step and the linearity of the distribution propagation on the lattice make the method highly suitable for its implementation on parallel computing architectures. This enables efficient utilization of modern multi-core processors and GPU-based systems.
In this study, we extend a recently developed high-order thread-safe version of the lattice Boltzmann method (TSLB) \cite{montessori2023thread, montessori2024order}, augmented with an interface capturing equation for the simulation incompressible, axisymmetric two-component turbulent jets at a unitary density ratio \cite{montessori2024high}. Specifically, we introduce a hybrid version of the TSLB capable of handling bi-component systems with high density and viscosity contrasts. This version leverages a thread-safe implementation optimized for shared memory architectures, such as those found in GPU-based devices. Additionally, the non-equilibrium part of the distribution function is locally reconstructed using the recursive properties of Hermite polynomials. This approach allows for the explicit inclusion of non-equilibrium hydrodynamic moments up to the third order—the highest order supported by the D3Q27 lattice—yielding significant improvements in both the stability and accuracy of the solver \cite{montessori2024high,malaspinas2015increasing}.

\section{Method}

\subsection{Navier-Stokes equations for multiphase flows with interfaces}

The multiphase system under investigation is modeled by solving the continuity and momentum equations using the lattice Boltzmann method. These equations read as follows:

\begin{equation}
    \partial_t \rho + \partial_\alpha (\rho u_\alpha) = 0,
\end{equation}

\begin{equation} \label{NSeq}
     (\partial_t (\rho u_\alpha) +  \partial_\beta( \rho u_\alpha u_\beta) ) = -\partial_\alpha p + \partial_\beta \big(\rho \nu [\partial_\beta u_\alpha + \partial_\alpha u_\beta]\big) + F^s_\alpha,
\end{equation}

where \( u_\alpha \) is the fluid velocity, \( \rho \) the density, \( p \) the macroscopic pressure, \( \nu \) the kinematic viscosity and \( F^s_\alpha \) the surface tension force. Greek indexes denote cartesian components of vectors and tensors. 

The surface tension force, \( F^s_\alpha \), can generally be expressed as the divergence of a capillary stress tensor:  
\[
F^s_\alpha = \partial_\beta \big(\sigma (\delta_{\alpha\beta} - n_\alpha n_\beta)\big),
\]
where \( n_\alpha \) is the local normal to the interface and $\sigma$ is the surface tension coefficient. In this work, we adopt the definition by Jaqmin et al.~\cite{jacqmin1999calculation}, where \( F^s_\alpha \) is expressed as:  

\begin{equation} \label{jaqminforce}
    F^s_\alpha = \mu_\phi \partial_\alpha \phi,
\end{equation}

with \( \mu_\phi \) representing the chemical potential for binary fluids (defined later) and \( \phi \) the local phase field.  

\subsubsection{ Allen-Cahn equation for interface tracking}

The interface tracking is performed by capturing the dynamic evolution of a phase field , i.e. $\phi$, via the following conservative Allen-Cahn equation:

\begin{equation}
    \partial_t\phi + u_\alpha \partial_\alpha \phi = D \partial_\alpha\partial_\alpha \phi - \kappa \partial_\alpha (\phi(1-\phi)n_\alpha)
\end{equation}

where $D$ is the diffusivity of the interface and $\kappa=4 D/\delta$ ($\delta$ being the interface width). 
In this model, $\phi$ can assume any value in the range $\{ 0,1 \}$, so that the interface between the two immiscible fluids is located at $\phi_0=0.5$.
It can be proven \cite{kim2005continuous} that the equilibrium profile for an interface located at $x_0$ is:

\begin{equation}
    \phi(x, y, z)=\phi_0 \pm \frac{\phi_H - \phi_L}{2}tanh\Bigg( \frac{|x-x_0|+|y-y_0|+|z-z_0|}{\delta}\Bigg)
\end{equation}

The interface tracking equation is dynamically coupled to the NSE equations through the advection term. Since the time-stepping in the LBM requires very small values of $\Delta t$, we chose to discretize the phase field, governed by the Allen-Cahn equation, using a forward-time, centered-space (FTCS) scheme. In this context, the use of the FTCS scheme for the Allen-Cahn equation has proven effective to maintain stability and consistency with the small time step constraints imposed by the LB framework, while efficiently handling the phase field dynamics. 
Given the above, we deemed a finite difference approach an appropriate choice for handling the Allen-Cahn equation, especially with the aim of reducing equation the overall computational complexity and memory footprint.
Future works will aim at investigating the effect of higher order discretization of the discrete derivatives along with the possibility to decouple the resolution of the Allen-Cahn and the momentum equation by solving the transport equation for the phase field on finer grids while maintaining coarser discretization of the momentum field \cite{schenk2024computationally,ding2014diffuse}.

\subsection{High-order thread-safe Lattice Boltzmann model}

The set of single-fluid Navier-Stokes equations introduce before can be solved via the lattice Boltzmann method. In this section, we provide a concise overview of the key components of the thread-safe LB framework.  This framework integrates a recently developed efficient LB strategy \cite{montessori2023thread} with a high-order version of the regularized LB, designed to utilize the recursive properties of Hermite polynomials \cite{grad1949kinetic,coreixas2017recursive,malaspinas2015increasing} to reconstruct non-equilibrium moments up to the third order.

To this aim, we start from the Boltzmann equation discretized over a lattice stencil of $q$ vectors in the velocity space:
\begin{equation} \label{lbe}
    f_i(x_\alpha+c_{i\alpha}\Delta t,t + \Delta t)=f_i(x_\alpha,t) + \omega(f^{eq}_{i}(x_\alpha,t)-f_i(x_\alpha,t)) + S_i(x_\alpha,t)
\end{equation}
where $x_\alpha$ and $t$ are the lattice position and time step, respectively, and $f_i(x_\alpha,t)$ is a set of distribution functions, being $i=1,..,q$ an index spanning the $q$ discrete velocity vectors $c_{i\alpha}$ of the lattice stencil. In Eq. \ref{lbe}, $f_i^{eq}$ represents a set of discrete thermodynamic equilibrium distributions derived from a Mach number expansion of the continuous Maxwell-Boltzmann distribution over the velocity space of $q$ discrete velocity vectors and $S_i(x_\alpha,t)$ is the \textit{i-th} component of an external forcing term, whose mathematical form will be fully defined in the following.  Finally, $\omega$ is a frequency tuning the relaxation towards the equilibrium distribution. If we denote the right hand side  of Eq. \ref{lbe} as $f^{post}_{i}$ indicating the post-collision distribution, it is easy to recognize that $f^{post}_{i}$ can be expressed as a weighted sum of the equilibrium and non-equilibrium parts of the $i^{th}$ probability distribution function (PDF) since we can always decompose the PDF as $f_i=f_i^{eq} + f_i^{neq}$. Thus, the post-collision PDF can written as:
\begin{equation}\label{LBexp}
    f^{post}_{i} = f^{eq}_{i} + (1-\omega)f^{neq}_{i} + S_i
\end{equation}

and, Eq. \ref{lbe} can be compactly written as:

\begin{equation}\label{pushLB}
    f_i(x_\alpha+c_{i\alpha}\Delta t,t + \Delta t) = f^{eq}_{i}(x_\alpha,t) + (1-\omega)f^{neq}_{i}(x_\alpha,t) + S_i
\end{equation}

The core of the thread-safe paradigm is that both the equilibrium and non-equilibrium parts can be reconstructed from the hydrodynamics macroscopic fields without the necessity to read PDFs, by resorting to a regularization process. 
Therefore, by reconstructing $f^{eq}_{i}$ and $f^{neq}_{i}$ at each time step, we can remove the data dependencies that occur during non-local read and write operations, thereby preventing race condition issues. The discrete set of equilibria in Eq.\ref{pushLB} reads as follows:

\begin{align}\label{eqts}
    f_i^{eq} = w_i \Bigg( p^* + \frac{c_{i\alpha} u_\alpha}{c_s^2} + \frac{(c_{i\alpha}c_{i\beta} - c_s^2 \delta_{\alpha\beta}) u_\alpha u_\beta}{2 c_s^4} \notag \\
    + \frac{(c_{i\alpha}c_{i\beta}c_{i\gamma} - c_{i\gamma}c_s^2 \delta_{\alpha\beta} - c_{i\alpha}c_s^2 \delta_{\beta\gamma} - c_{i\beta}c_s^2 \delta_{\alpha\gamma}) u_\alpha u_\beta u_\gamma}{6 c_s^6} \Bigg)
\end{align}
where $p*$ is a non dimensional pressure defined as $p*=p/(\rho c_s^2)$ and $p$ is the local pressure.
The non-equilibrium part of the PDF can be reconstructed up to $n=3$ as \cite{malaspinas2015increasing,montessori2024high}:
\begin{equation} \label{noneqts}
    f_i^{neq}=\frac{c_{i\alpha}c_{i\beta} a^2_{neq,\alpha\beta}}{2 c_s^4} +  \frac{(c_{i\alpha}c_{i\beta}c_{i\gamma}-c_{i\gamma}c_s^2\delta_{\alpha\beta} - c_{i\alpha}c_s^2\delta_{\beta\gamma} - c_{i\beta}c_s^2\delta_{\alpha\gamma}) a^3_{{neq},\alpha\beta\gamma}}{6 c_s^6},
\end{equation}
where $a^3_{neq,\alpha\beta\gamma}=u_\alpha a^2_{{neq},\beta\gamma}+u_\beta a^2_{{neq},\alpha\gamma}+u_\gamma a^2_{{neq},\alpha\beta}$ using the recursive relation of Hermite polynomials\cite{grad1949note}, and $a_{{neq},\alpha\beta}^{2}=-\frac{1}{\omega c_s^2} (\partial_{\alpha}\rho u_\beta+\partial_{\beta}\rho u_\alpha)$.

The interested reader is referred to \cite{malaspinas2015increasing} where a detailed description on the derivation of both the high order equilibrium and non-equilibrium part is provided.

Once the fused stream-and-collision step (Eq.\ref{pushLB}) is performed, the updated values of the macroscopic moments  are retrieved via the following computation of the relevant statistical moments:
\begin{equation} \label{moments0}
    p^* (x_\alpha,t)=\sum_i f_i(x_\alpha,t),
\end{equation}
\begin{equation} \label{moments1}
    u(x_\alpha,t)=\sum_i f_i(x_\alpha,t)c_{i\alpha} + \frac{1}{2}\sum_i S_i
\end{equation}
and the components of the non-equilibrium second order tensor, $a_{neq}$, which are linked to the deviatoric part of the stress tensor through:
\begin{equation} \label{hermite_neq2}
    a_{neq,\alpha \beta}^2=\sum_i (c_{i\alpha}c_{i_\beta}-c_s^2\delta_{\alpha\beta}) (f_i- \overline{f_i^{eq}}).
\end{equation}
$\overline{f_i^{eq}}=f_i^{eq} - \frac{1}{2} S_i$ and $S_i$ is encoded in the collision step as follows \cite{guo2002discrete}:

\begin{equation} \label{guoforce}
    S_i=w_i \left( \frac{c_{i\alpha} - u_\alpha}{c_s^2} + \frac{c_{i\beta} u_\beta}{c_s^4}c_{i\alpha}\right)F_\alpha ,
\end{equation}

being $F_\alpha$ the Cartesian components of the force field. Such a redefinition of the equilibrium distribution function, along with the use of the forcing scheme in Eq.\ref{guoforce}  is needed to implement the trapezoidal rule in the collision step when an external forcing term is applied \cite{kruger2017lattice}.

The sequence of operations in Eq. \ref{pushLB},\ref{moments0},\ref{moments1},\ref{hermite_neq2} is then repeated at each time iteration.

\subsection{Model extension for high-density and viscosity contrasts}

The above lattice Boltzmann equation must be extended to reproduce, at the macroscopic level, the set of Navier-Stokes equations as in Eq.\ref{NSeq}.
In particular, a set of forces must be introduced to model a positive surface tension between interacting fluid interfaces  and to take into account the additional pressure and viscous forces arising due to the inhomogeneity of the density field. 
Thus the vector $F_\alpha$ can be splitted into three contrbutions, namely:

\begin{equation}
    F_\alpha=F_\alpha^s + F_\alpha^p +F_\alpha^\nu, 
\end{equation}

As denoted in eq.\ref{jaqminforce} $F_\alpha^s=\mu_\phi\partial_\alpha\phi$ where $\mu_\phi=4\beta(\phi-\phi_l)(\phi-\phi_g)(\phi-\phi_0) - \kappa \partial_\alpha^2 \phi$, where $\beta=12\sigma/\delta$ and $\kappa=3\sigma\delta/2$.\\

The pressure force is
\begin{equation}\label{press_corr}
F_\alpha^p=-p^*c_s^2 \partial_\alpha \rho
\end{equation}
being $\rho=\rho_l \phi+(1-\phi)\rho_g$ and $\rho_l$ and $\rho_g$ the densities in the liquid and gas phase respectively. 

Indeed, by employing the equilibria as in Eq.\ref{eqts} the local pressure reads as $p=p^*\rho c_s^2$. 

Thus by computing the gradient and splitting the two contributions we obtain $\nabla p =\rho c_s^2 \nabla p^* + p^* c_s^2 \nabla \rho$. The first term is already embedded in the equation via the chosen form of the discrete equilibria, while the second one must be introduced via the external forcing term $F_\alpha^p$.

Lastly, the viscous force $F_\alpha^\nu=\nu\left( \partial_{\alpha}u_\beta+\partial_{\beta} u_\alpha \right) \partial_\alpha \rho$ can be implemented by exploiting the existing relation between  the second order moment of the discrete set of distributions and the deviatoric stress tensor as \cite{kruger2009shear}:

\begin{equation}\label{visc_corr}
    F_\alpha^\nu= -\frac{\nu\omega}{c_s^2\Delta t}\left[\sum_i (f_i - f_i^{eq})c_{i\alpha}c_{i\beta}\right] \partial_{\alpha} \rho
\end{equation}

In passing we note that, gradients and laplacians have been computed by employing the following discrete formulas \cite{THAMPI20131}:

\begin{equation}
    \partial_\alpha\Psi=\frac{1}{c_s^2}\sum_i w_i \Psi(x_\alpha + c_{i\alpha})c_{i\alpha} ,
\end{equation}

\begin{equation}
    \partial_\alpha\partial_\beta\Psi=\frac{1}{c_s^2}\left(\sum_{i\neq 0} w_i \Psi(x_\alpha + c_{i\alpha}) -w_0\Psi(x_\alpha) \right)
\end{equation}

being $\Psi$ a generic scalar or vector field.

\subsection{Implementation of the hybrid LB-finite difference implementation}

Summarizing, the main steps needed to implement of the hybrid LB-finite difference model for multiphase flows with high density and viscosity contrasts are the following:

\begin{itemize}
    \item \textbf{Moments computation:}
    \begin{itemize}
        \item For each \(i,j,k\), compute \(p^*\) as in Eq.~\ref{moments0}, the pressure correction force (Eq.~\ref{press_corr}), the surface tension force as in Eq.~\ref{jaqminforce}, and the correction to the viscous dissipation term as in Eq.~\ref{visc_corr}. 
        \item Compute the velocity components as in Eq.~\ref{moments1}.
    \end{itemize}

    \item \textbf{Pre-regularization:}
    \begin{itemize}
        \item For each \(i,j,k\), reconstruct the non-equilibrium second-order tensor as in Eq.~\ref{hermite_neq2}.
    \end{itemize}

    \item \textbf{TSLB (fused streaming-collision)}
    \begin{itemize}
        \item Perform the fused streaming and collision at each lattice node, via Eq.~\ref{LBexp}, by reconstructing the non-equilibrium set of distributions as in Eq.~\ref{noneqts}.
    \end{itemize}

    \item \textbf{Interface tracking:}
    \begin{itemize}
        \item Compute the local density from the phase field by recalling that $\rho=\phi\rho_L + (1-\phi)\rho_G$, $\rho_L$ and $\rho_G$ the liquid and gas density respectively, and the gradient and Laplacian of \(\phi\).
        \item Compute the local value of \(\phi\) at the subsequent time step through finite difference via:
        \begin{equation}
        \phi^{n+1} = \phi^{n} - (u_\alpha \partial_\alpha \phi)^n + D (\partial_\alpha \partial_\alpha \phi)^n - \kappa (\partial_\alpha(\phi(1-\phi) n_\alpha))^n
        \end{equation}
    \end{itemize}
    where $n$ is the $n-th$ time step
\end{itemize}

\subsection{Performances on Nvidia A100 GPU}

\begin{table}[h!]
\caption{The GLUPS, MPI decomposition along $x, y, z$ axis, speed up ($S_p$), and parallel efficiency ($E_p$) versus the number of computing GPU devices, ($n_{p}$),  for bi-component fluid simulation at high-density ratio (water-air) in strong scaling. Results are reported from the top to bottom for a cubic box of sides 512 and 1024. Note that the 1024-size simulations start from 4 GPUs due to the large GPU-memory requirements.}
\begin{tabular}{l l l l l}
{$n_{p}$} & {MPI decomp.} & GLUPS & {$S_p$} & {$E_p$} \\ \hline \hline 
1 & $1 \times 1 \times 1$ & 1.49 & 1.00 & 1.00 \\
2 & $1 \times 1 \times 2$ & 2.85 & 1.91 & 0.96 \\
4 & $1 \times 1 \times 4$ & 5.42 & 3.63 & 0.91 \\
8 & $2 \times 2 \times 2$ & 8.31 & 5.58 & 0.70 \\
16 & $2 \times 2 \times 4$ & 13.52 & 9.07 & 0.57 \\
32 & $2 \times 4 \times 4$ & 20.34 & 13.65 & 0.43 \\ \hline
4 & $1 \times 1 \times 4$ & 5.94 & 4.00 & 1.00 \\
8 & $2 \times 2 \times 2$ & 10.29 & 6.93 & 0.87 \\
16 & $2 \times 2 \times 4$ & 19.65 & 13.24 & 0.83 \\
32 & $2 \times 4 \times 4$ & 36.17 & 24.37 & 0.76 \\ \hline
\end{tabular}
\label{tab:performance-strong}
\end{table}

\begin{table}[h!]
\caption{The GLUPS, MPI decomposition along $x, y, z$ axis, speed up ($S_p$), and parallel efficiency ($E_p$) versus the number of computing GPU devices, ($n_{p}$),  for bi-component fluid simulation at high-density ratio (water-air) in weak scaling fixing a cubic sub-domain of side size equal to 512 lattice points for each GPU device. Note that $S_{p}$ has been redefined as a function of GLUPS in Eq. \ref{eq:speedup}. }
\begin{tabular}{l l l l l}
{$n_{p}$} & {MPI decomp.} & GLUPS & {$S_p$} & {$E_p$} \\ \hline \hline 
1 & $1 \times 1 \times 1$ & 1.49 & 1.00 & 1.00 \\
2 & $1 \times 1 \times 2$ & 2.93 & 1.96 & 0.98 \\
4 & $1 \times 1 \times 4$ & 5.85 & 3.91 & 0.97 \\
8 & $1 \times 1 \times 8$ & 11.50 & 7.70 & 0.96 \\
16 & $1 \times 1 \times 16$ & 22.96 & 15.37 & 0.96 \\
32 & $1 \times 1 \times 32$ & 45.34 & 30.35 & 0.95 \\ \hline
\end{tabular}
\label{tab:performance-weak}
\end{table}

\begin{figure}
    \centering
    \includegraphics[width=0.8\linewidth]{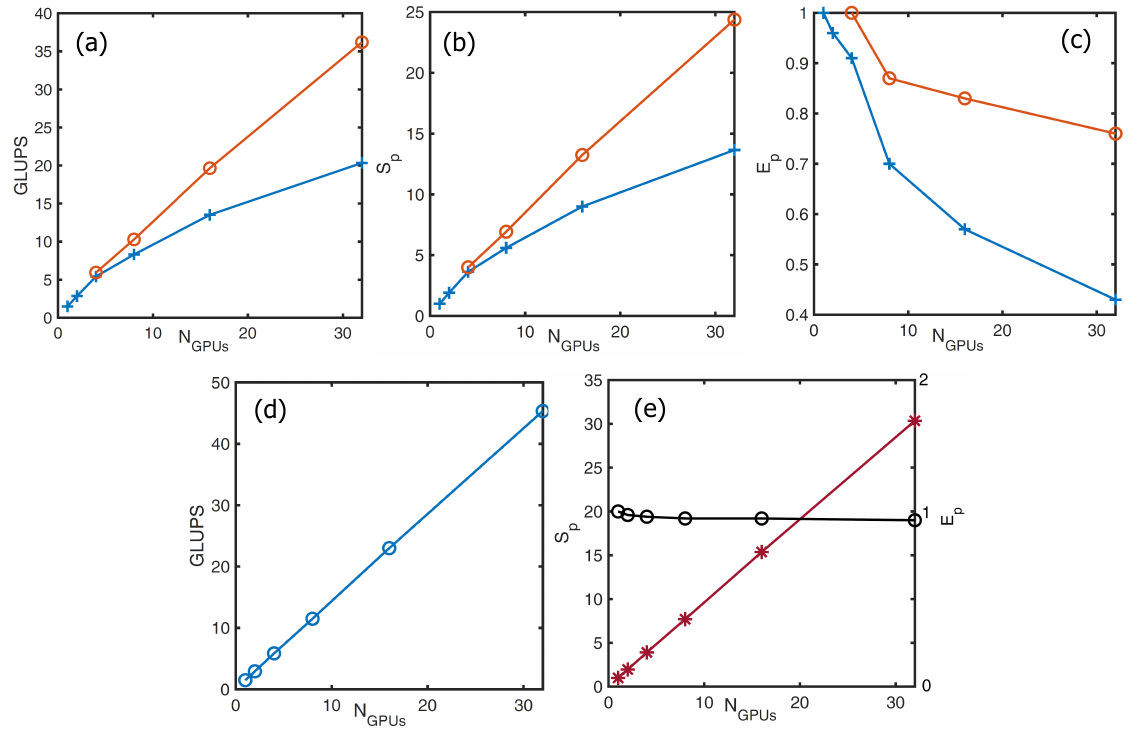}
    \caption{Plot of the strong scalings for cubic boxes of sides $512$(panels (a) to (c) blue lines with plusses) and $1024$ (panels (a) to (c) red lines with circles), as reported in TABLE I. Panels (d-e) report the weak scaling for data reported in TABLE II. Note that $S_{p}$ has been redefined as a function of GLUPS in Eq.\ref{eq:speedup}. In panel (e) the line with asterisks represents $S_p$ while the one with hollow circles $E_p$.}
    \label{performances}
\end{figure}

We probe the efficiency of the implemented parallel strategy by quantitative estimators (see Fig.\ref{performances} and Tables I and II).
To measure the performance, the Giga Lattice Updates Per Second (GLUPS) unit is used. In particular, 
the definition of GLUPS reads:
\begin{equation}
\label{eq:glups}
\text{GLUPS}=\frac{L_x L_y L_z}{10^9 t_{\text{s}}},
\end{equation}
where $L_x$, $L_y$, and $L_z$ are the domain sizes in the
$x-$, $y-$, and $z-$ axis, and $t_{\text{s}}$ is the run (wall-clock) time (in seconds)
per single-time step iteration.

Additionally, we introduce the speedup ($S_{p}$), which is redefined in the present work as:

\begin{equation}
S_{p}=\frac{\text{GLUPS}_{p}}{\text{GLUPS}_{s}},
\label{eq:speedup}
\end{equation}

where $\text{GLUPS}_{s}$ represents the Giga Lattice Updates Per Second for the code running on a single GPU, used as the baseline, and $\text{GLUPS}_{p}$ denotes the Giga Lattice Updates Per Second for the code running in parallel mode on $n_{p}$ GPU cards.

Hence, the parallel efficiency ($E_{p}$) is expressed as
\begin{equation}
E_{p}=\frac{S_{p}}{n_{p}}.
\end{equation}

The benchmarks were performed on a GPU-based HPC cluster (LEONARDO, a petascale supercomputer located at the CINECA data center in Italy). In particular, each node is equipped with a single socket 32 cores Intel Xeon Platinum 8358, Ice Lake CPU,
and four NVIDIA Ampere GPUs/node, 64GB HBM2e connected by NVLink 3.0 (200GB/s). The inter-node communications are managed by a DragonFly+ NVIDIA Mellanox Infiniband HDR (25 GB/s). 
In all the following benchmarks, the source code was compiled
using the Nvidia Compiler, version 24.3 while 
the MPI library was exploited to manage parallel communications.
Hence, strong scaling results are reported in Tab. \ref{tab:performance-strong} for two fixed cubic box sizes equal to 512 and 1024 lattice points over different domain decompositions (variable sub-domain size).
The benchmark results clearly show that the effect of parallel communications is mitigated by the size of the computational domain. To highlight the effect of parallel communications, we report the weak scale, where the sub-domain size is fixed for each MPI process (GPU device). In Tab. \ref{tab:performance-weak}, we show the weak scaling fixing a cubic sub-domain of side size equal to 512 lattice points for each GPU device (fixed sub-domain size). Thus, a weak scaling is performed by adopting a linear decomposition along the z-axis, showing remarkable results in terms of parallel efficiency $E_p$.

In an era where machine learning (ML) often seems indispensable for computational investigations, we point out that the exceptional performance attainable with modern GPU-based supercomputers makes ML non-essential for studying complex flows, even in the presence of fluid interfaces. The present results highlight the feasibility of conducting systematic, high-resolution simulations of highly complex multiphase flows without ML assistance. This suggests that state-of-the-art computational resources and highly optimized algorithms are sufficient to resolve such systems with a fidelity that would be challenging for ML-based methods to achieve. 

\section{Results}

To evaluate the capability of the multiphase TSLB model in handling complex multiphase flows with significant density and viscosity contrasts, we conducted two sets of benchmark simulations. The first one focused on the rising motion of a bubble in a heavy fluid, while the second one analyzed both head-on and off-axis collisions of two water droplets in air.

As a potential application, we extended our analysis to simulate the dynamics of multiple droplets impacting a solid substrate. This simulation is particularly relevant for investigating the behavior of raindrops striking the ground, with implications for assessing the ejection and dispersion of microplastics and other environmental pollutants.

All the simulations were performed on the Leonardo supercomputer using the multi-GPU open-source code \textit{accLB} \cite{montessori2024high, montessori2024order}. The code was executed in parallel on four Nvidia A100 GPUs, corresponding to a single computational node of the Leonardo system. This setup enabled high computational efficiency and the ability to capture the intricate physics of the multiphase flow phenomena under scrutiny.

\subsection{Rising bubble in a quiescent liquid environment}

In this subsection, the rising dynamics of a gas bubble evolving in a liquid quiescent environment is investigated. The motion of the bubble is obtained via a body force is applied exclusively to the light dispersed phase.
The phenomenon is governed by the Eötvös (\( Eo \)) and Galilei (\( Ga \)) numbers, defined as:

\begin{equation}
  Eo = \frac{\rho_L g R^2}{\sigma},
\end{equation}
\begin{equation}
  Ga = \frac{\sqrt{g R} \cdot R}{\nu_L},
\end{equation}

where \( g \) is the gravitational acceleration, \( R \) the bubble radius, \( \sigma \) the surface tension coefficient, and \( \nu_L \) and \( \rho_L \) the kinematic viscosity and density of the liquid phase, respectively. The density and viscosity ratios of the system are set to 1000 and 100, reflecting the properties of an air-water system. 

\begin{figure}
    \centering
    \includegraphics[width=0.5\linewidth]{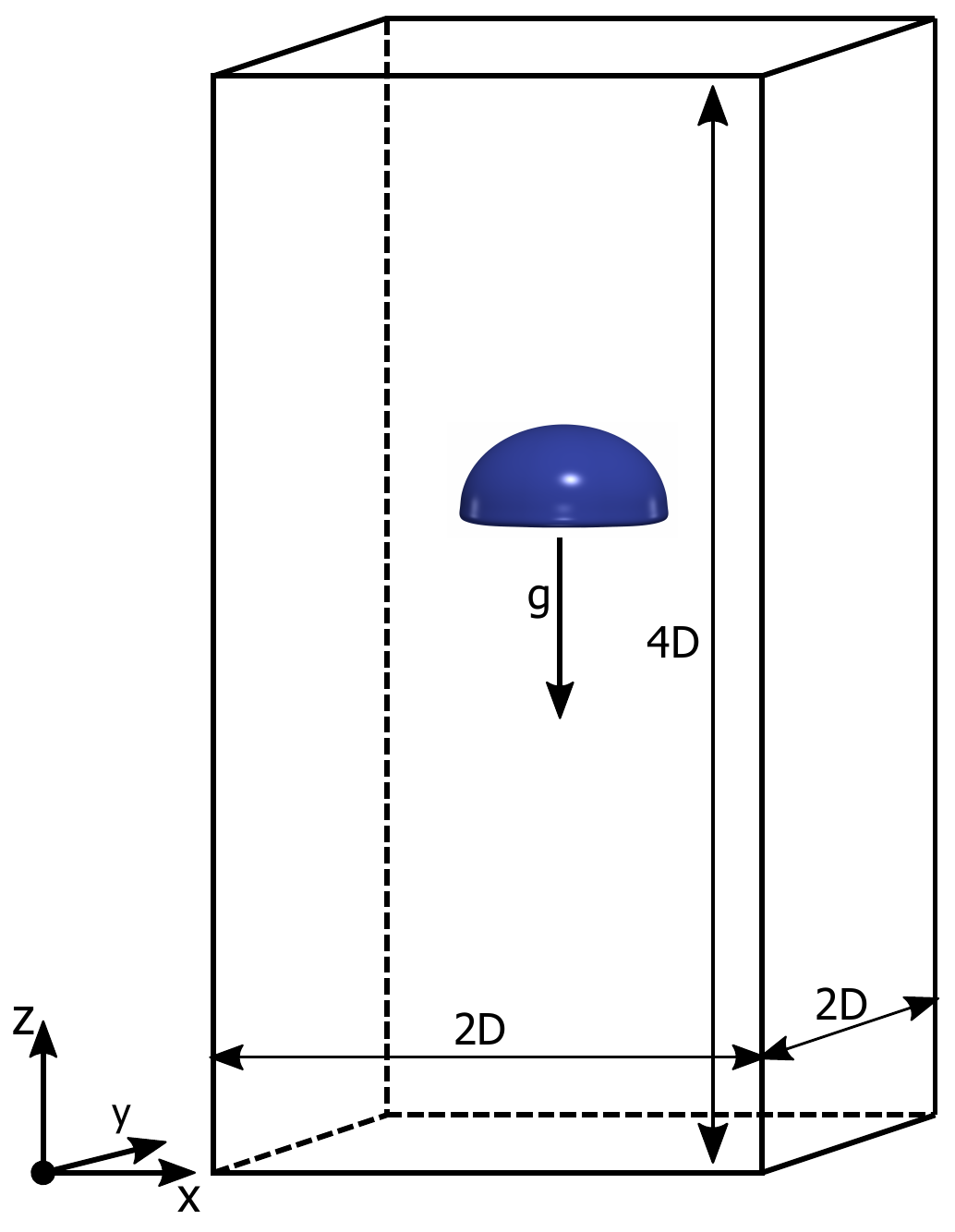}
    \caption{Sketch of the computational domain for the rising bubble test. $D$ is the bubble diameter set to $100$ lattice units in all the simulations. The domains counts $200\times200\times400$ grid points.}
    \label{sketchbub}
\end{figure}

\begin{figure}
    \centering
    \includegraphics[width=1\linewidth]{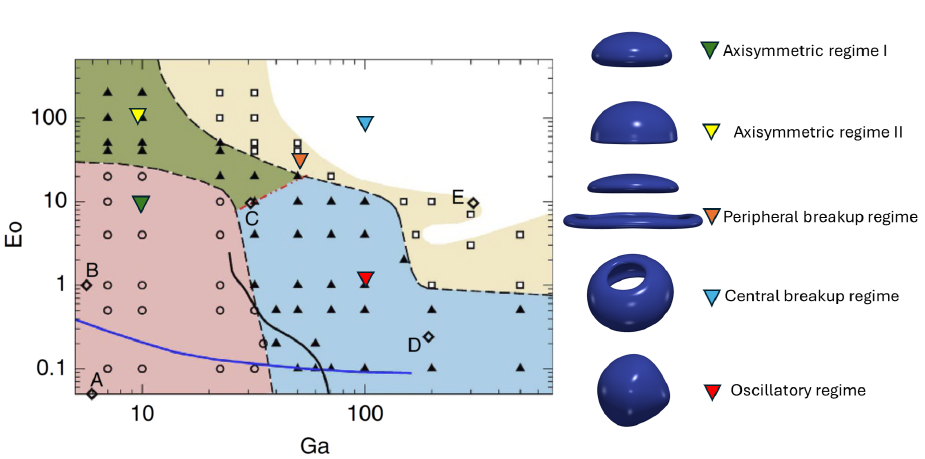}
    \caption{Deformation regimes of a rising bubble as a function of \( Eo \) and \( Ga \).}
    \label{fig:eogamap}
\end{figure}

In this first set of simulations, periodic boundary conditions are enforced in all directions. The size of the domain is reported in figure \ref{sketchbub} and Fig. \ref{fig:eogamap}  summarizes the deformation regimes of the rising bubble for various \( Eo \) and \( Ga \) values \cite{tripathi2015dynamics}.

The multiphase TSLB model captures five distinct deformation modes on the Eo-Ga map. These include the axisymmetric regimes: \textbf{Regime I} (small deformation) and \textbf{Regime II} (hat-shaped deformation), both characterized by steady, axisymmetric deformations. 

For \( Eo = 30 \) and \( Ga = 100 \), the model enters the \textbf{peripheral breakup regime}. Initially, the bubble deforms like in Axisymmetric Regime II. However, higher inertial forces lead to the detachment of a peripheral gas toroid that subsequently fragments into smaller bubbles.

\begin{figure}
    \centering
    \includegraphics[width=0.55\linewidth]{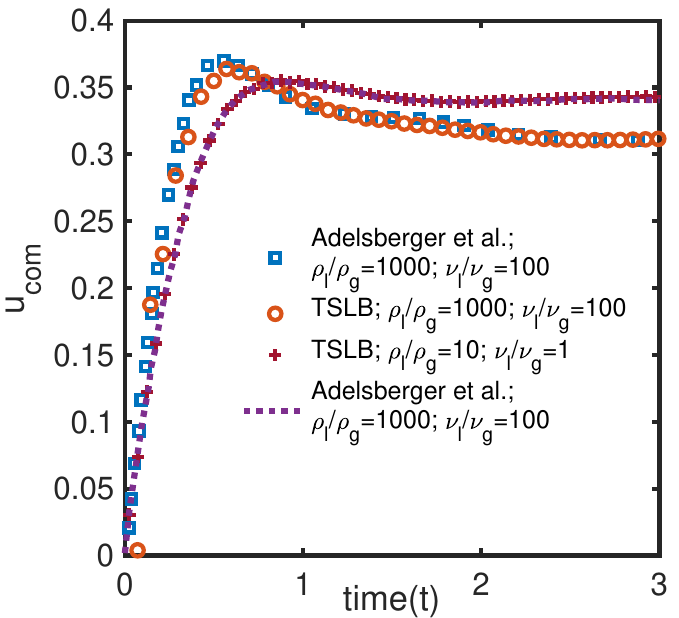}
    \caption{Center-of-mass velocity ($u_{com}$) over time for the simulated cases.}
    \label{fig:comparison}
\end{figure}

At higher values of \( Eo \) and \( Ga \) (approximately \( Eo > 10 \), \( Ga > 100 \)), inertial forces dominate over viscous and surface tension ones, resulting in central breakup during ascent. The bubble, initially spherical, rises due to buoyancy. As it ascends, its lower surface collapses due to the onset of an upward liquid jet, triggered by a pressure difference between the upper and the lower surface of the gas bubble. This jet pushes gas upward, forming a toroidal protrusion at the top of the bubble, as depicted in Fig. \ref{fig:comparison}.

The bottom-right region of the Eo-Ga diagram corresponds to the \textbf{oscillatory regime}. Here, low \( Eo \) values and high \( Ga \) values indicate that surface tension resists bubble deformation, while inertial forces overcome viscous dissipation, inducing hydrodynamic instabilities in the wake. As the bubble rises, wake vortices start oscillating, breaking flow symmetry and causing the bubble to randomly move in the liquid medium.

To further validate the TSLB model's accuracy in simulating buoyant multiphase flows, we perform two sets of simulations: one with \( Eo =10 \) and \( Ga =35 \) for \( \rho_L/\rho_G = 10 \), \( \nu_L/\nu_G = 1 \), and another with \( Eo =125 \), \( Ga =35 \), \( \rho_L/\rho_G = 1000 \), and \( \nu_L/\nu_G = 100 \) (i.e., air-water-like system). No-slip boundary conditions are applied on each wall, as in  \cite{adelsberger20143d} . Figure \ref{fig:comparison} compares the velocity of the bubble's center of mass with reference results from \cite{adelsberger20143d}, showing good agreement provided that the interface width is adequately resolved, in particular in the high density ratio case. The simulations confirm that the TSLB model accurately reproduces the rising dynamics even for the challenging air-water case.

\subsection{Head-on collision between water droplets in air}

We proceed by comparing  lattice Boltzmann simulations and experimental data from \cite{ashgriz1990coalescence} on head-on impacts of water droplets in air. The impacts were investigated across a range of Weber ($We=\rho_LU^2D/\sigma$) and Reynolds numbers($Re=UD/\nu$). Specifically, the Weber number ranged from 25 to 96, while the Reynolds number varied between $1100$ and $2100$, to match the experimental dimensionless numbers. In this way the Ohnesorge number ($Oh= \sqrt{We/Re}$) is kept constant throughout both experiments and simulations, $Oh=0.0044$. 
\begin{figure}
    \centering
    \includegraphics[width=0.5\linewidth]{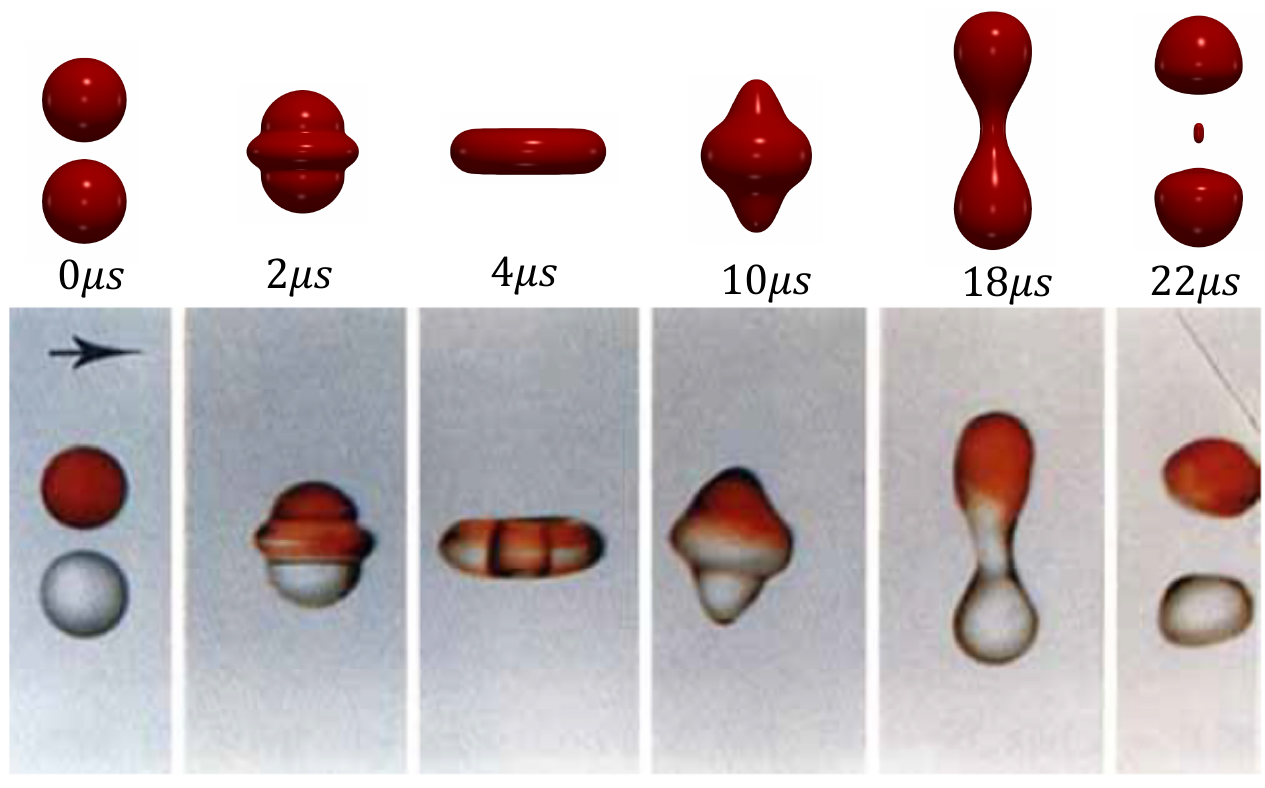}
    \caption{Reflexive head-on separation at $We = 23$ and $Re=1136$. Comparison between simulation (upper panel) and experiments(lower panel) \cite{ashgriz1990coalescence}.}
    \label{We23Re1136}
\end{figure}

\begin{figure}
    \centering
    \includegraphics[width=0.5\linewidth]{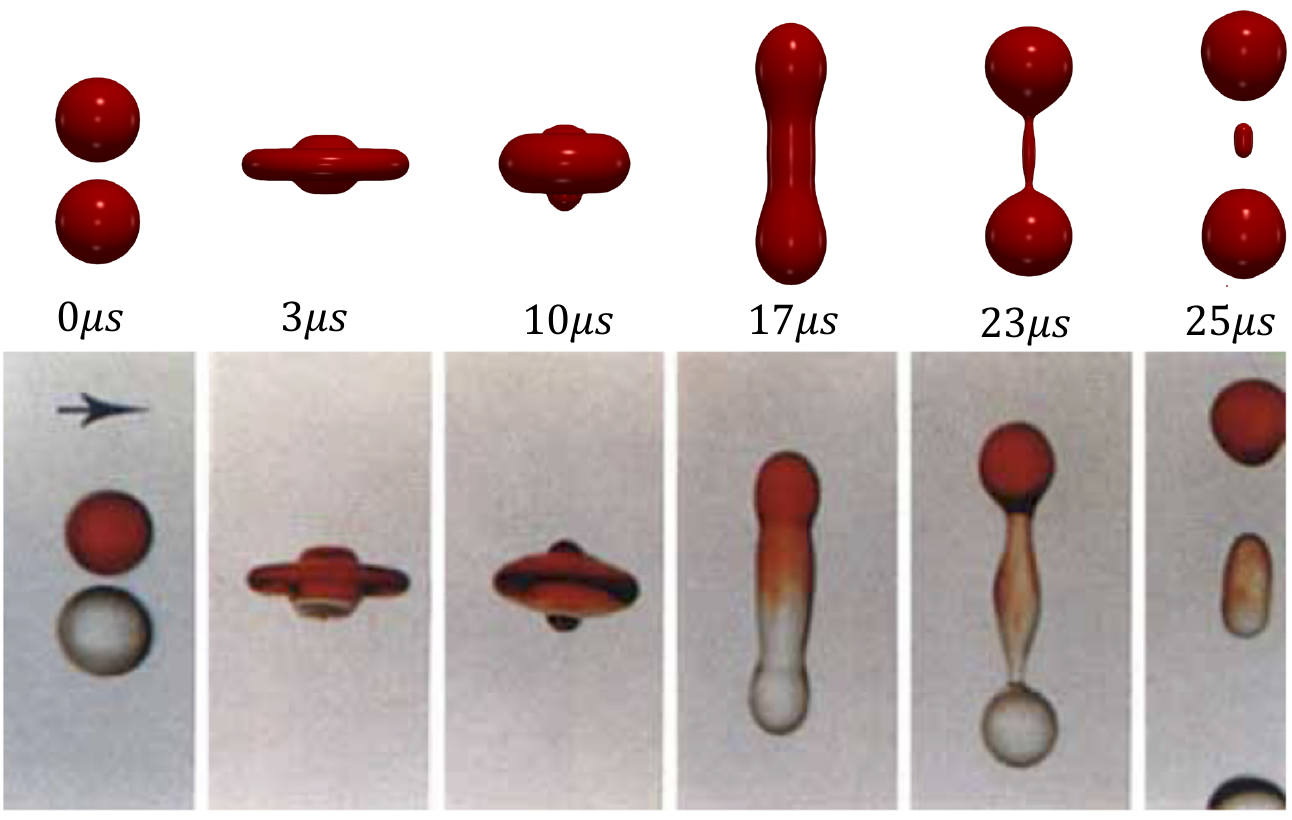}
    \caption{Reflexive head-on separation with satellite droplet formation at $We = 40$ and $Re=1440$. Comparison between simulation (upper panel) and experiments(lower panel) \cite{ashgriz1990coalescence}.}
    \label{We40Re1140}
\end{figure}

\begin{figure}
    \centering
    \includegraphics[width=0.35\linewidth]{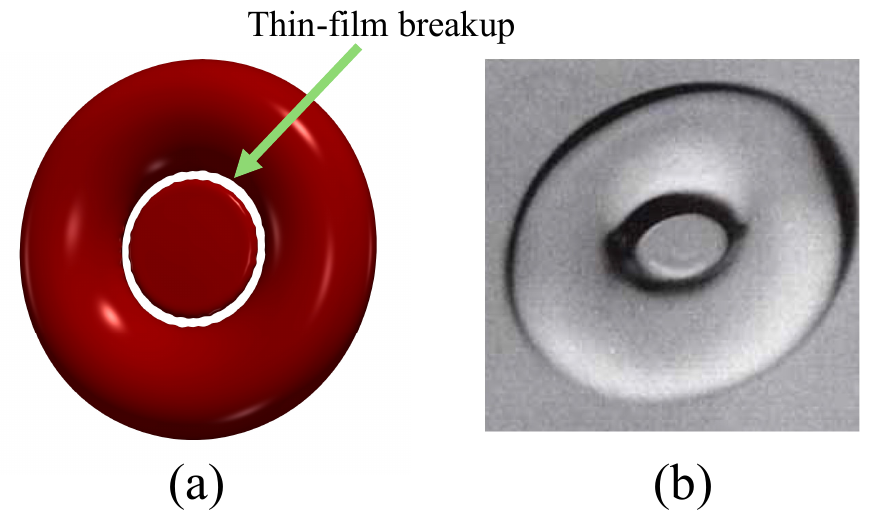}
    \caption{Top view of head-on collision at $We=40$. Comparison between simulation (a) and experiments (b)\cite{ashgriz1990coalescence}. At sufficient high Weber number the discretization is insufficient to prevent the rupture of the thin film in the connecting the fluid torus with the internal thin fluid sheet. }
    \label{thinfilmbreak}
\end{figure}

\begin{figure}
    \centering
    \includegraphics[width=0.4\linewidth]{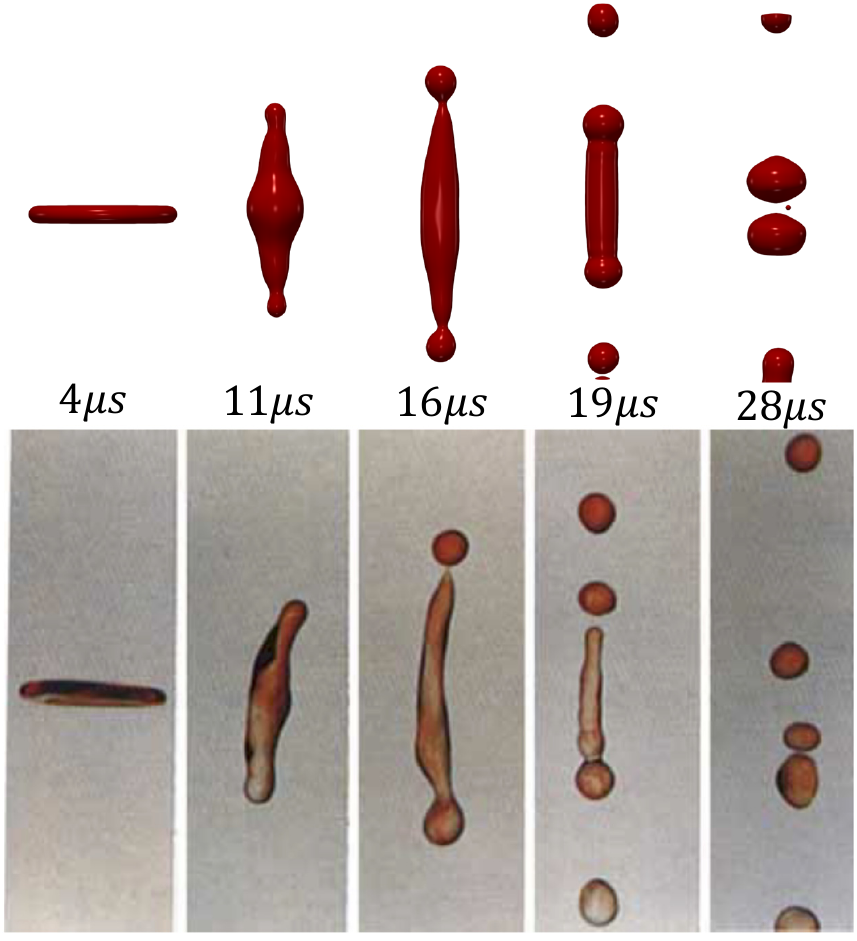}
    \caption{Reflexive head-on separation with formation of multiple satellite droplets at $We = 96$ and $Re=2230$. Comparison between simulation (upper panel) and experiments (lower panel) \cite{ashgriz1990coalescence}.}
    \label{We96Re2230}
\end{figure}

The simulations were conducted using a grid size of $ 500 \times 500 \times 700$ lattice nodes.

The diameter of the droplets has been discretized with $100$ lattice units, and the interface thickness set to 4 lattice units, yielding a Cahn number  (namely the ratio between interface width and the droplet diameter) of $Ca=0.04$. This value is sufficient to accurately capture the dynamic behavior of the droplets during the impact process \cite{magaletti2013sharp}. The density ratio and the viscosities of the two fluids have been chosen in order to match the experimental values and set to  $830$ and $15$, respectively. In the simulations the Weber has been varied by changing the value of the surface tension, while the Reynolds number has been matched by varying the viscosities of the droplet and continuous phase. The resulting impact velocity has been kept fixed at $u_{impact}=0.005\;lu/step$. Since the experiments reported in \cite{ashgriz1990coalescence} do not provide clear information on the exact droplet size, we set the spatial resolution to $\delta x=0.5\mu m$, resulting in a droplet diameter of $50$ microns. Using these values, we applied dynamic scaling to compute the impact times reported in the figures.

In Figure \ref{We23Re1136}, snapshots from a simulation of reflexive head-on droplet separation are presented and compared graphically with the experiments of \cite{ashgriz1990coalescence}. As observed, the droplets approach each other at a relative velocity of $U_{\text{impact}}$, collide, and merge into two hemispheres separated by a toroidal fluid structure at $t = 2 \, \mu s$. This torus increases in size until it reaches maximum elongation at $t = 4 \, \mu s$. Due to the high curvature at the circumference of this disk-torus-shaped drop, a pressure difference increases between its inner and outer regions. Consequently, the disk contracts radially inward, expelling liquid from its center. This contraction is thus a reflexive action of the liquid surface. The reflexive motion eventually generates a long cylindrical structure with rounded caps at $t = 18 \, \mu s$. If the Weber number is sufficiently high, as in this case, a critical condition is reached, causing the liquid cylinder to break into at least two droplets.

At $We = 25$, experimental observations from \cite{ashgriz1990coalescence} do not report the formation of satellite droplets. However, our simulations show the formation of a very small droplet after the breakup of the liquid thread connecting the separating droplets. Interestingly, recent experimental observations by \cite{huang2019pinching} report the formation of at least one droplet following the pinch-off of the thread between the colliding droplets. This behavior aligns with the universal features of a thinning liquid filament approaching singularity, as predicted by scaling theories of pinch-off. The absence of satellite droplet formation at low Weber numbers can be attributed to minor symmetry-breaking during the impact.

When the Weber number is increased to $We = 40$ (Fig. \ref{We40Re1140}), a larger satellite droplet forms after the pinching of the cylindrical thread, consistent with experimental observations. The satellite droplet predicted by the simulation is slightly smaller than that observed in \cite{ashgriz1990coalescence}, likely due to the formation of a fluid sheet detached from the main the toroidal structure during the early stages of impact (between $3 \, \mu s$ and $10 \, \mu s$), an effect not observed in the experiments. As shown in Fig. \ref{thinfilmbreak}, the breakup of the connecting thin film may reduce the amount of elastic energy stored during the elongation process, influencing the overall dynamics of the reflexive separation between the droplets. It should be noted that the formation of such a hole in the thin film of the toroidal structure is likely because of the insufficient resolution used to capture thin film dynamics, as mentioned in \cite{amani2019numerical}. Nonetheless the simulation keeps on providing a physically-sound outcome of the head-on collision process.

\subsection{Off-axis collision between water droplets in air}

To further assess the predictive capabilities of the multiphase TSLB model, we conducted a comprehensive set of simulations to analyze off-axis impacts of water droplets in air. 
\begin{figure}
    \centering
    \includegraphics[width=0.55\linewidth]{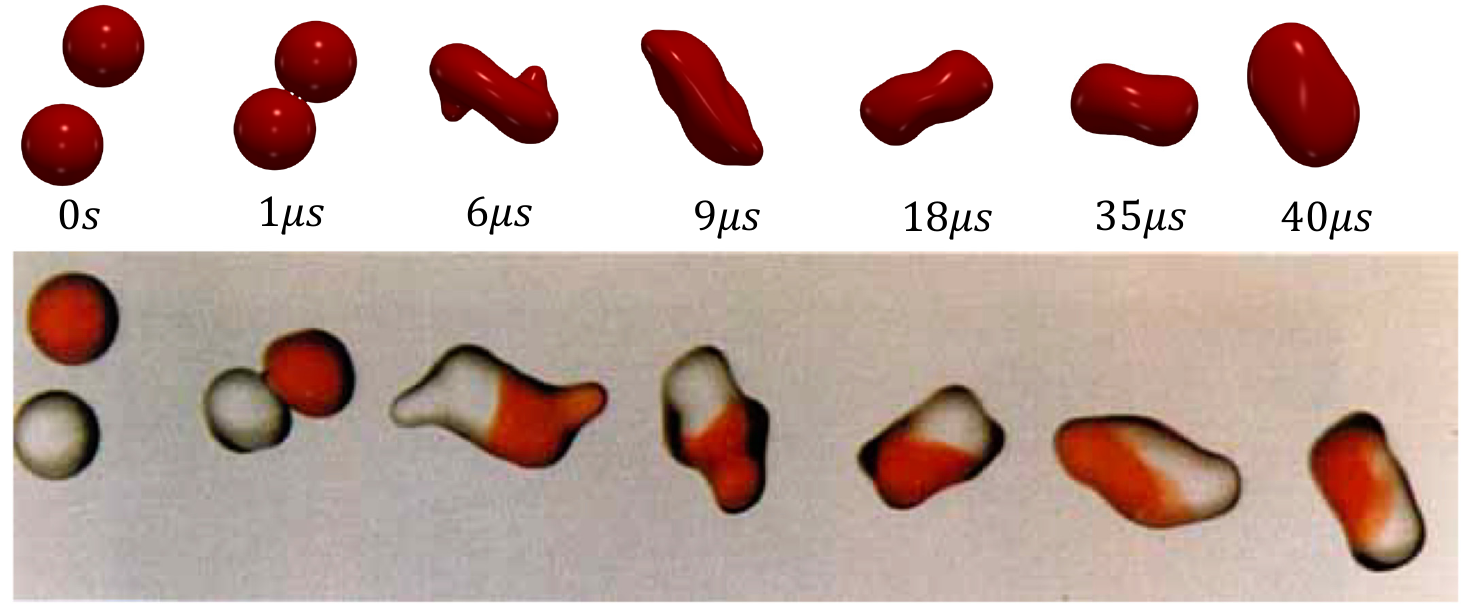}
    \caption{Coalescence collision at $We = 10$, and x = 0.5. Comparison between simulation
(upper panel) and experiments(lower panel) \cite{ashgriz1990coalescence}. }
    \label{We10x5}
\end{figure}

\begin{figure}
    \centering
    \includegraphics[width=0.55\linewidth]{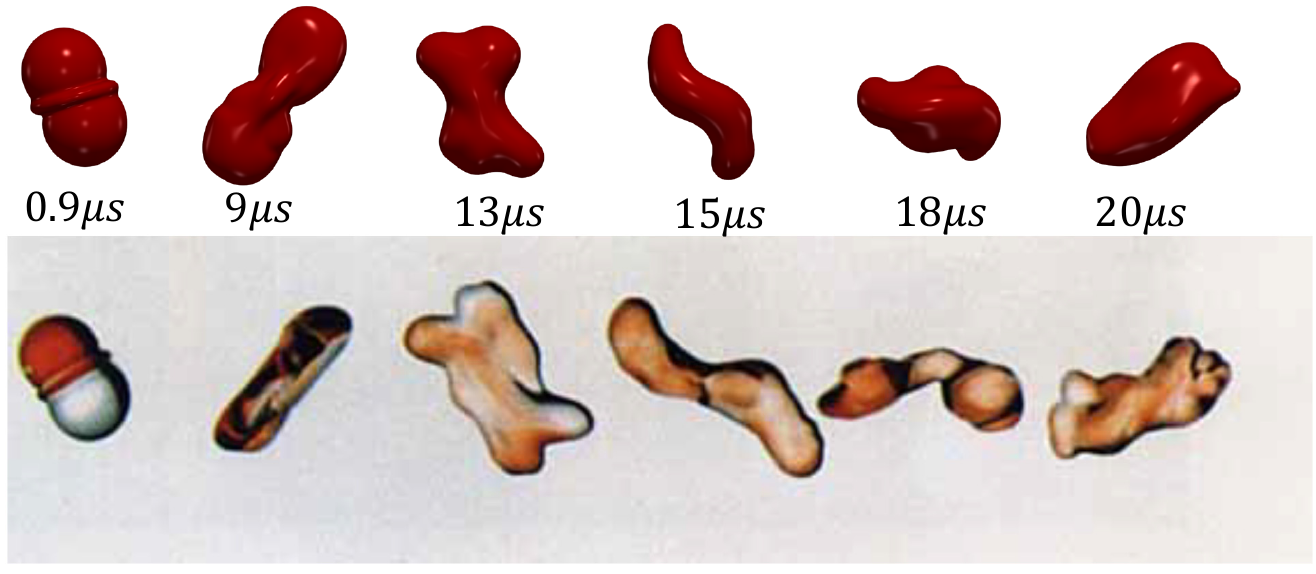}
    \caption{Coalescence collision at $We = 53$, and x = 0.28. Comparison between simulation
(upper panel) and experiments(lower panel) \cite{ashgriz1990coalescence}. }
    \label{We53x28}
\end{figure}

\begin{figure}
    \centering
    \includegraphics[width=0.5\linewidth]{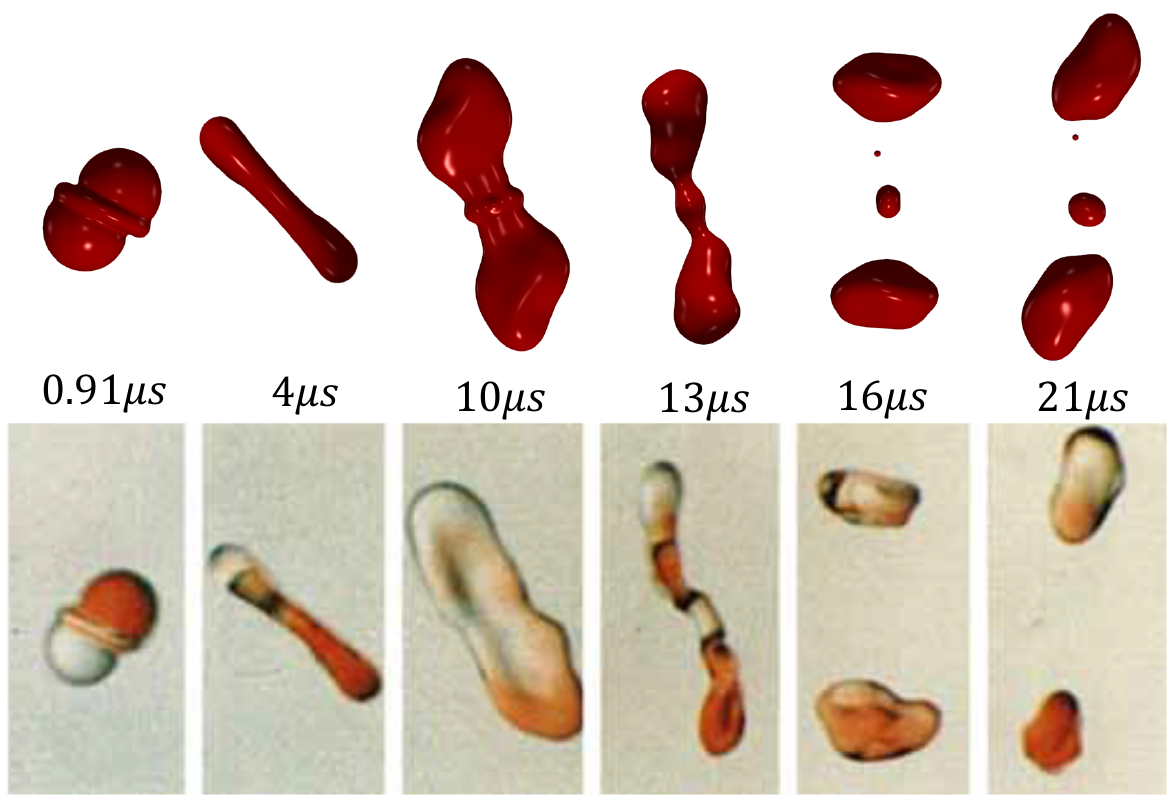}
    \caption{Collision with satellite droplet formation of at $We=53$ and $x=0.38$. Comparison between simulation
(upper panel) and experiments(lower panel)\cite{ashgriz1990coalescence}.}
    \label{We53x38}
\end{figure}

\begin{figure}
    \centering
    \includegraphics[width=0.5\linewidth]{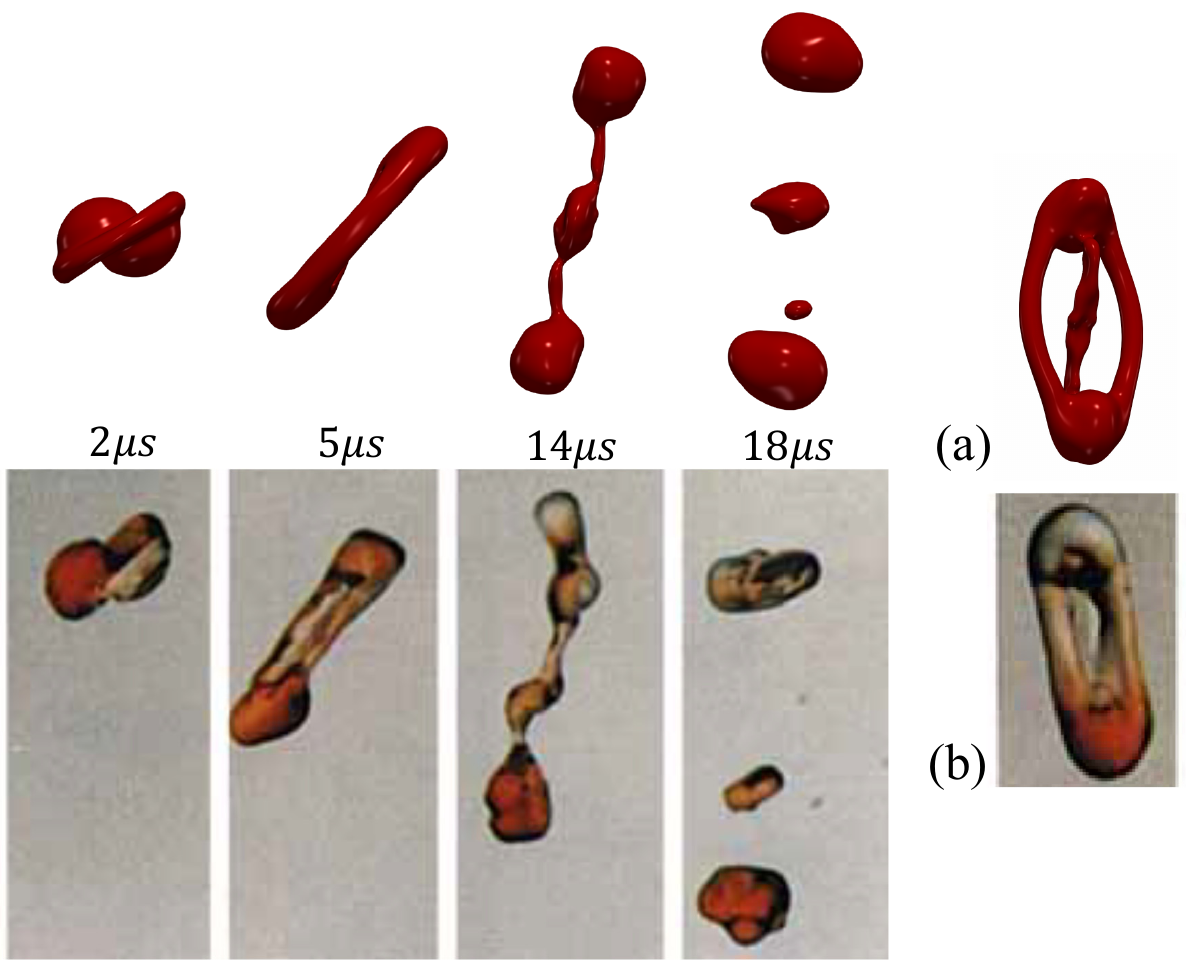}
    \caption{Collision with satellite droplet formation of at $We=83$ and $x=0.38$.Comparison between simulation
(upper panel) and experiments(lower panel)\cite{ashgriz1990coalescence}.}
    \label{We83x34}
\end{figure}

\begin{figure}
    \centering
    \includegraphics[width=0.43\linewidth]{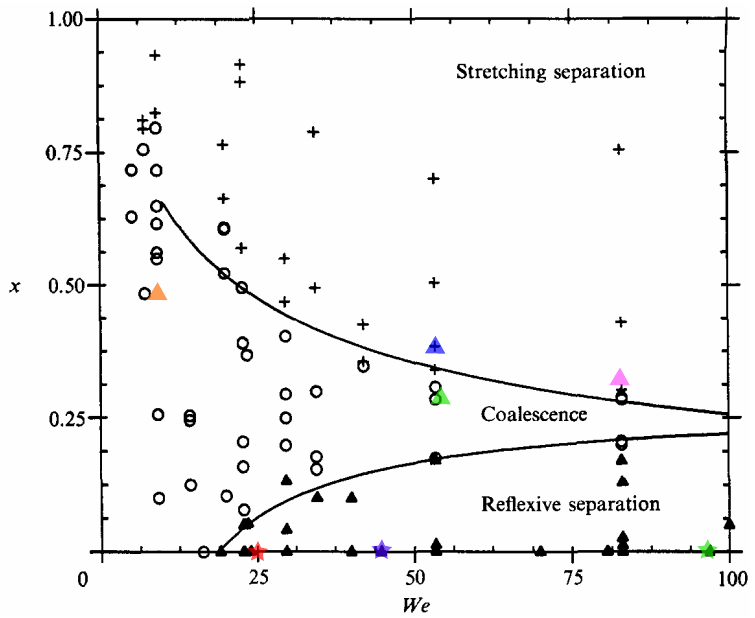}
    \caption{$x-We$ plot reporting the map of collision parameter and dynamic conditions leading to different collision outcomes. Coloured triangles represent the numerical simulations performed in this work.}
    \label{xwe}
\end{figure}
The results  were systematically compared with experimental data reported in \cite{ashgriz1990coalescence}. The investigation focused on capturing the impact dynamics over a broad range of Weber ($We$) and Reynolds ($Re$) numbers. Specifically, the Weber number was varied from 10 to 83, while the Reynolds number ranged from 720 to 2100, ensuring alignment with the dimensionless parameters used in the reference experiments. To maintain consistency across simulations, the $Oh$ was kept constant at $0.0044$.

Fig. \ref{We10x5} and \ref{We53x28} illustrate two representative cases at relatively low Weber numbers, corresponding to $x=0.5$ and $We=10$, and $x=0.28$ and $We=53$, respectively. $x$ is impact parameter, i.e. the ratio of the distance between the centers of the droplets and their diameter. In these cases, the droplets undergo collision and rotation subject to a torque. The combined action of surface forces and inertia prevents separation, leading to the coalescence of the two droplets into a single merged entity. This behavior aligns with experimental observations, where coalescence is expected under similar impact conditions.

As the Weber number increases, the collision dynamics become significantly more complex. Higher values introduce larger inertial forces, resulting in more pronounced deformation of the droplets. This deformation can lead to stretching and separation of the colliding droplets, with the possibility of forming satellite droplets depending on the dynamic conditions and collision parameters. Figures \ref{We53x38} and \ref{We83x34} showcase two cases of off-axis collisions, with $x=0.38$ and $We=53$, and $x=0.34$ and $We=83$, respectively. The multiphase TSLB model successfully predicts the outcomes of these collisions, demonstrating close agreement with experimental observations. In these scenarios, the model accurately captures key features such as droplet elongation, interface breakup, and the subsequent formation of smaller satellite droplets.

Furthermore, as summarized in Fig.\ref{xwe}, which presents a collision map in the $x-We$ plane, the TSLB model exhibits exceptional flexibility in tuning relevant physical parameters. By adjusting surface tension, density, and viscosity ratios, the model is capable of reproducing a wide range of collision regimes, from coalescence to separation. This adaptability allows the model to accurately capture the transition between these regimes, mirroring the behavior observed in the experimental results from \cite{ashgriz1990coalescence}. Overall, the predictive power and adaptability of the TSLB model make it a valuable tool for investigating the intricate dynamics of off-axis droplet collisions in air.

\subsection{Multiple raindrops collisions onto a solid surface}

We report here a simulation of the dynamics of rain droplets impacting a solid substrate. 

The simulation setup consists of a train of $5$ droplets falling under the effect of an initial velocity and the gravity. The domain is periodic east-west, rear-front and no-slip boundary conditions have been enforced on the bottom and top wall via the thread-safe boundary condition \cite{montessori2024order}. As per the phase field, the boundary conditions are periodic at each side rather than on top and bottom walls  Robin boundary conditions are enforced to code for impermeable, neutral wetting walls. The simulation domains counts $500\times500\times 3000$ nodes.

\begin{figure}
    \centering
    \includegraphics[width=0.65\linewidth]{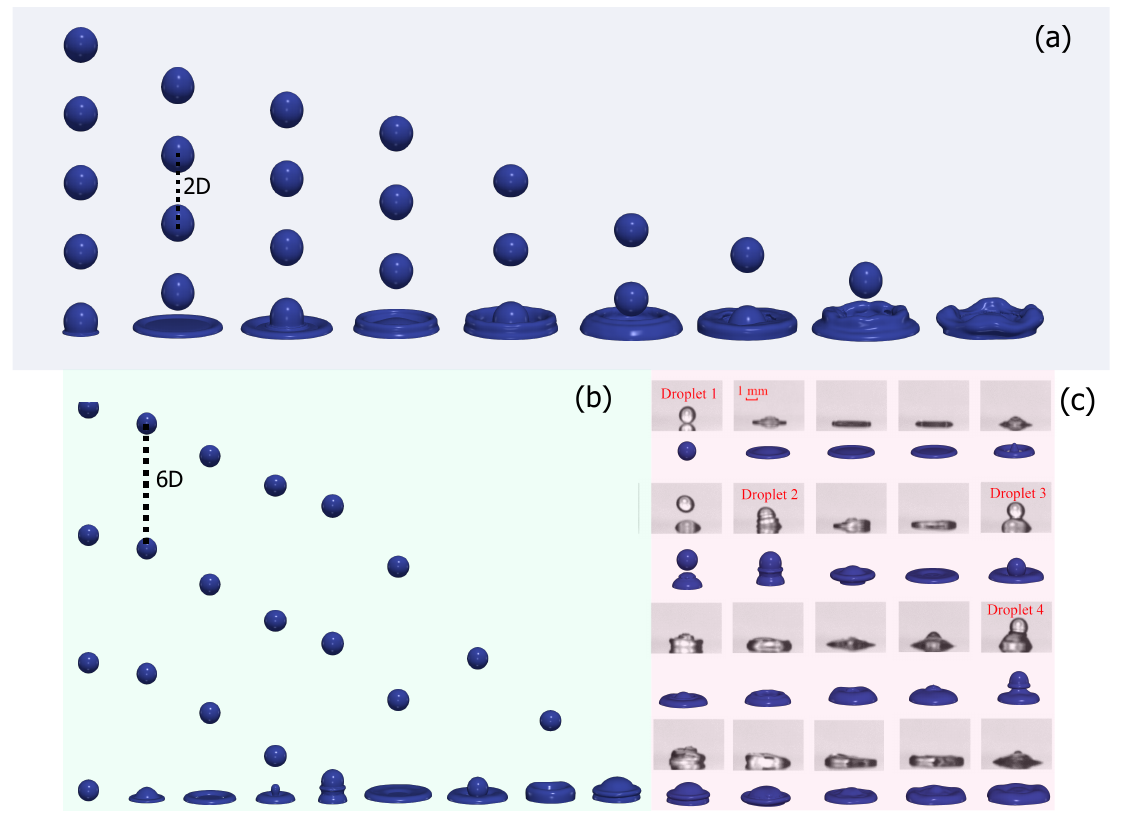}
    \caption{(a-b) sequence of multiple impacts between falling droplets and a neutral solid substrate at $We\sim40$ with different distances between subsequent falling droplets($2D$ and $6D$). (c)Visual comparison  between simulation and experiments\cite{li2019dynamics} of a train of droplets impacting a solid surface at $We \sim 40$ and $\Delta L=6D$ being $\Delta L$ the distance between the centres of the impacting dorplets.The simulation was conducted with a higher impact frequency (i.e., smaller distance between centers of subsequent impacting droplets) compared to the experimental conditions.Despite this difference, the main dynamic features are accurately captured by the multiphase TSLB model.
    }
    \label{fig:rain_droplet_impact}
\end{figure}

The simulation considers rain droplets falling at a Weber number $We = 40$ and the distance between the centers of two subsequent droplets being $\Delta L= 2$ and $\Delta L=6$ diameters(see Fig.\ref{fig:rain_droplet_impact}). The second configuration   is  similar to one of the  experimental conditions reported in \cite{li2019dynamics}, in which the distance between two subsequent inpacting droplets is roughly $\sim
8$ diameters. As can be seen from Fig. \ref{fig:rain_droplet_impact}, the physical behavior observed in the simulation  aligns with the experimental observations. This similarity underscores the accuracy and predictive capability of the multiphase LB model in reproducing the complex dynamics of droplet impact and fragmentation.

The figure illustrates the evolution of the droplet impact process at different time instants. Upon contact with the solid substrate, the droplet undergoes significant deformation, leading to the formation of thin liquid sheets followed by an quasi-elastic retraction due to the relatively high value of surface tension  over inertial forces. 

When the first droplet makes contact with the surface, it deforms into a truncated sphere. The bottom of the droplet compresses, forming a thin liquid lamella that spreads radially from the point of collision. As the spread diameter of the liquid film increases, the height of the film decreases rapidly. During this process, the kinetic energy of the droplet is converted into surface energy. When the spread diameter reaches its maximum value, the rim of the lamella swells slightly, marking the moment when the surface energy also peaks. Following this, the outer edge of the lamella gradually moves back toward the center under the influence of capillary forces. Simultaneously, the liquid film merges, causing the height at the center of the film to rise.

In the case of a droplet train impact, the deforming liquid film on the surface is impacted by subsequent droplets before the oscillating impinged droplet reaches its equilibrium state. During each subsequent impact, the falling droplet does not approach the solid surface directly but instead interacts with the hemispherical liquid film formed by the preceding droplet. Upon contact, the falling droplet and the liquid film coalesce at the contact point and spread radially in the plane normal to the contact point. This process generates a circular lamella that moves outward and downward due to inertial forces and gravity, eventually coalescing with the surface.

Following the second impact, the liquid film extends beyond its equilibrium position, leading to a recoil and re-spread similar to the behavior observed after the first impact. However, the maximum spread diameter and the film height during the deformation are greater than those observed during the initial impact. For subsequent impacts, the deformation process of the liquid film on the surface repeats. As the cumulative mass of liquid on the surface increases, both the spread diameter and the film height increase with each successive impact.

Overall, the results of this simulation demonstrate the potential of the multiphase LB model to reproduce and analyze the complex fluid dynamics of rainfall impact. This capability is essential for advancing our understanding of pollutant dispersion mechanisms in natural water systems, particularly in the context of microplastic atmospheric pollution. Future studies could extend this approach to investigate a wider range of impact conditions, including different Weber numbers, surface properties, and environmental factors, to gain a comprehensive understanding of the factors that influence pollutant spread during rainfall events.  
\section{Conclusions}

In this work we presented a high-order, thread-safe version of the lattice Boltzmann method, incorporating an interface-capturing equation, based on the Allen-Cahn equation, to simulate incompressible two-component systems with high density and viscosity contrasts. The method relies on
a recently proposed, high-order thread-safe implementation optimized for shared memory architectures and it
is employed to reproduce the dynamics of droplets and bubbles on two benchmark cases, namely the rising bubble in a quiescent liquid and head-on and off-axis collision between equally-sized droplets in air. The results are in agreement with experiments and other numerical simulations from the literature. In particular the model is capable of reproducing correct dynamic behavior of bubbles and droplets in air-water like systems (i.e., with density and viscosity ratios as high as $1000$ and $100$ respectively). As a perspective application we presented some preliminary results of a train of raindrops impacting onto a solid surface. This type of simulation is particularly relevant for the detailed investigation of the behavior of rain droplets striking the ground with practical importance for assessing the ejection and dispersion of microplastics in the environment. The proposed approach offers promising opportunities for high-performance computing simulations of realistic
systems, particularly on GPU-based architectures.
\label{sec:sample1}

\section*{Acknowledgements}

A.M. and M.L acknowledge fundings from the Italian Government through the PRIN (Progetti di Rilevante Interesse Nazionale) Grant (MOBIOS) ID: 2022N4ZNH3 -CUP: F53C24001000006 and computational support of CINECA through the ISCRA B project DODECA (
HP10BUBFIL). S.S., M.L. and A.T. acknowledge the support from the European Research Council under the ERCPoC Grant No. 101187935 (LBFAST).
M.L. and A.T. acknowledge the support of the Italian National Group for Mathematical Physics (GNFM-INdAM).

 \bibliographystyle{elsarticle-num} 





\end{document}